\documentclass[aps,prl,twocolumn,groupedaddress,amsmath,amssymb]{revtex4}

\usepackage{graphicx}
\usepackage{dcolumn}
\usepackage{bm}

\begin{document}
\title{Dephasing of a superconducting flux qubit}
\author{K. Kakuyanagi$^{1}$, T. Meno$^{2}$, S. Saito$^{1}$,
H. Nakano$^{1}$,
K. Semba$^{1}$, H. Takayanagi$^{3}$,
F. Deppe$^{4}$, and A. Shnirman$^{5}$}
\affiliation{
$^{1}$NTT Basic Research Laboratories, NTT Corporation, Kanagawa, 243-0198, Japan\\
$^{2}$NTT Advanced Technology, NTT Corporation, Kanagawa, 243-0198, Japan\\
$^{3}$Tokyo University of Science, 1-3 Kagurazaka, Shinjuku, Tokyo 162-8601, Japan\\
$^{4}$Walther-Mei\ss ner-Institut,Walther-Mei\ss ner-Str. 8, D-85748 Garching, Germany \\
$^{5}$Institut f\"{u}r Theoretische Festk\"{o}rperphysik, Universit\"{a}t Karlsruhe, D-76128 Karlsruhe, Germany}

\begin{abstract}
In order to 
gain a better understanding of the origin of
decoherence in superconducting flux qubits, we have measured the magnetic 
field dependence of the characteristic 
energy relaxation time ($T_1$) 
and echo phase relaxation time ($T_2^{\rm echo}$) near the optimal operating point of a flux qubit.
We have measured $T_2^{\rm echo}$ by means of the phase cycling method.
At the optimal point, we found the relation $T_2^{\rm echo}\approx 2T_1$.
This means that the echo decay time is 
{\it limited by the energy relaxation }
($T_1$ process) 
.
Moving away from the optimal point, 
we observe a 
{\it linear }
increase of the phase relaxation rate ($1/T_{2}^{\rm echo}$) with the applied external magnetic flux.
This behavior 
can be well explained by the influence of magnetic flux noise with a $1/f$ spectrum on the qubit.
\end{abstract}


\maketitle

As a consequence of the progress in the micro-fabrication technology, 
in the last years solid-state based 
artificial quantum two level systems (TLS), which can be 
fabricated, operated and read out in a controlled way have been successfully realized \cite{SCQ2,SCQ4}.
These artificial 
TLS 
can be used as quantum bits (qubits), i.e. the fundamental devices for quantum information systems \cite{QC}.

Among the solid state based qubits one of the most promising candidates are the Josephson junction based qubits ~\cite{SCQ2,SCQ4,SCQ5}.
Because superconductivity is a macroscopic quantum phenomenon,
superconducting qubits have an inherent advantage over qubits based on microscopic systems such as 
single atoms~\cite{MSS}.
Furthermore, Josephson junction based qubits also represent promising candidates for the study of fundamental aspects of macroscopic quantum systems
 \cite{LG0}.
For these 
studies
, long coherence time as well as high readout 
fidelity 
 are required.
Evidently, the 
fidelity 
can be improved by refining the detection method, such as Josephson bifurcation amplifier technique \cite{JBA}.
However, decoherence is still determined by the environment of the qubit.
Thus, in order to reduce decoherence we have to first understand its origin in detail.

In this letter, we present a systematic study of the dephasing in a superconducting flux qubit whose level splitting is sensitive to the applied external magnetic flux.
This allows us to investigate the dependence of the coherence time on the applied flux bias.
It is expected that such a measurement gives information on
the interdependence between the qubit decoherence and the magnetic fluctuations.
We present experimental results on the magnetic flux dependence of both the energy relaxation time ($T_1$) and the echo phase decay time ($T_2^{\rm echo}$) of a superconducting flux qubit.
Based on our data, we discuss possible mechanisms responsible for decoherence in superconducting flux qubits.
As a result, we find that at the optimal point the coherence time
is limited
 by the energy relaxation ($T_1$ process).
Away from the optimal point, the phase relaxation time is determined by $1/f$ magnetic fluctuations.

\begin{figure}[t]
\begin{center}
\includegraphics[width=0.9\columnwidth]{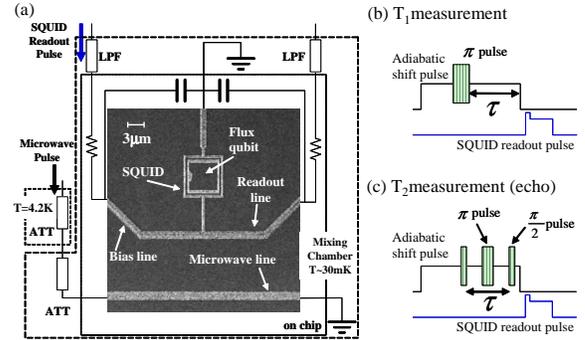}
\caption{(color online). (a) Scanning electron micrograph of the sample
.
Three Josephson junction flux qubit is surrounded with a SQUID detector loop.
We apply 
MW
 and adiabatic dc shift pulse via the 
MW-line.
The mutual inductance between the 
MW-line
 and the flux qubit is $\sim 1$ pH.
The 
readout pulse
is applied via the current bias line 
and 
qubit 
state is detected by measuring the voltage through the readout line. We use copper powder filter as low-pass filter (LPF) of these lines. (b) Pulse sequence for the $T_1$ measurement. We generate composite pulse by adding adiabatic DC pulse and 
MW
 pulse. 
The dc-SQUID readout pulse is programmed to reach the sample just after the adiabatic shift pulse is turned off. A rising time of each 
MW
 pulse is faster than 0.5 ns. The  $\pi$ pulse width is about $\sim$2 ns. (c) Pulse sequence for the $T_2^{\rm echo}$ measurement. In order to avoid a reduction of the visibility by the $T_1$ process, we programmed the end of the adiabatic shift pulse and the start of the readout pulse to be reached to the sample as soon as the second $\pi/2$ pulse finished.
}
\label{PULSE}
\end{center}
\end{figure}

The sample 
consists of a three junction flux qubit which is surrounded by a dc SQUID
 \cite{PARAM}.
Unlike in other qubit designs~\cite{SCQ4},
the SQUID loop does not share a current line with the qubit loop.
Thus the coupling between qubit and readout SQUID is purely inductive
(see Fig.~\ref{PULSE}(a)) 
with the mutual inductance 
$\sim 13$ pH.
The electromagnetic environment of the qubit
is engineered by equipping the SQUID detector with an on-chip shunting capacitor.
For the fabrication of the Al-Al$_2$O$_3$-Al Josephson junctions we used usual angled shadow evaporation technique~\cite{READOUT}.
The sample was mounted in a dilution refrigerator and cooled it down below 50mK.

At the optimal flux bias point, the mean value of the persistent current vanishes for both states of the qubit.
To be able to discriminate between the two states of the qubit at this operation point, we used the adiabatic shift pulse method as illustrated in Fig.~\ref{PULSE}(b,c)~\cite{V-Rabi}.
Then we apply an adiabatic magnetic dc-pulse with a rise time of 0.8 ns to the qubit through the coplanar RF-line. The qubit operation is performed during this adiabatic shift pulse. At the end of the shift pulse, the qubit will move back to the initial bias point adiabatically, where the readout can be done.
In this way we can get information about the qubit state even close to the optimal point.

\begin{figure}[b]
\begin{center}
\includegraphics[width=0.75\columnwidth]{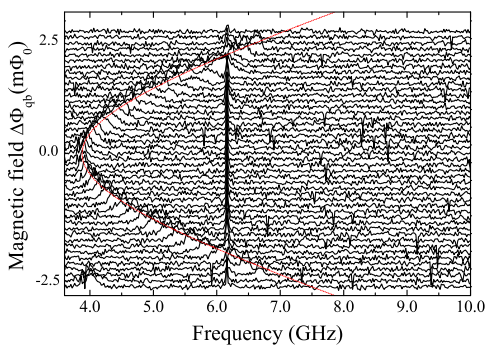}
\caption{(color online). Magnetic flux dependence of frequency sweep spectrum.
The qubit spectrum is observed as a hyperbola.
The field insensitive resonance at 6.2 GHz is attributed to the on-chip LC resonator~\cite{V-Rabi}. The red dotted line represents  a fit of the qubit spectrum by a hyperbolic function.
}
\label{SPECTRUM}
\end{center}
\end{figure}

We first performed 
MW
 spectroscopy of the qubit.
The resonant frequency (corresponding to the energy splitting of the qubit) is determined by saturating the state of the qubit with a long enough 
MW
 pulse and measuring the SQUID switching probability afterwards.
Figure.~\ref{SPECTRUM} shows the result obtained for the magnetic flux dependence of the level splitting of the qubit. 
In the two-level approximation the qubit Hamiltonian is $H_{qb} = \frac{1}{2} \left(\epsilon(\Phi_x)\sigma_z + \Delta \sigma_x\right)$, where $\epsilon(\Phi_x)$ is the level splitting induced by the external flux $\Phi_x$ threading the qubit loop and $\sigma_x$, $\sigma_z$ are the Pauli matrices.
From the spectroscopy experiment we determined the qubit gap frequency (i.e. the energy splitting at the optimal point) to be 
$\Delta/h=3.9$ GHz, where $h$ is the Planck constant. We also found the persistent current away from the optimal point, 
${h \over 2}{\partial \epsilon \over \partial \Phi_x} = I_p=370$ nA.

Next, we measured the magnetic flux dependence of $T_1$ and $T_2^{\rm echo}$.
$T_1$ is the energy relaxation time describing the transition of the qubit from the excited state to the ground state due to the interaction with the environment (for $k_{\rm B} T\ll \Delta$).
In order to measure $T_1$, we use the 
pulse sequence shown in Fig.~\ref{PULSE}(b).
Repeating this sequence provides the relaxation probability as a function of the waiting time.
$T_1$ is obtained by an exponential fit of this curve.
The $T_2^{\rm echo}$ time is the decay time in 
the echo experiments
depicted in Fig.~\ref{PULSE}(c).

To understand the effect of non-ideal control in the echo sequence we calculate the state of the qubit after applied three pulse
sequence in the absence of fluctuations.
The integrals (rotation angles) of the three pulses are $\theta_1, \theta_2$ and
$\theta_3$.
The time interval between the first and the second pulses is $\tau_1$, while between the second and the third pulse it is $\tau_2$.
We assume that the initial qubit state is along the z-axis of the Bloch sphere and calculate the z-axis projection of the final state $\left<\sigma_z\right>$:
\begin{eqnarray}
\label{ECHO}
\left< \sigma_z\right> &=&\cos{\theta_1} \cos{\theta_2} \cos{\theta_3}\nonumber
-\sin{\theta_1} \sin{\theta_2} \cos{\theta_3} \cos{\delta\omega\tau_1}\nonumber\\
&&-\cos{\theta_1}\sin{\theta_2}\cos{\theta_3} \cos{\delta\omega\tau_2}\nonumber\\
&&-\sin{\theta_1}\left( \cos{\theta_2}+1\right)\sin{\theta_3}
\cos{\left(\delta\omega\left(\tau_1+\tau_2\right)\right)}/2\nonumber\\
&&-\sin{\theta_1}\left(
\cos{\theta_2}-1\right)\sin{\theta_3}
\cos{\left(\delta\omega\left(\tau_1-\tau_2\right)\right)}/2\ ,
\end{eqnarray}
Here, $\delta\omega$ is the detuning.
For an ideal echo sequence we have $\theta_1 = \theta_3=\pi/2$, $\theta_2 = \pi$, and $\tau_1=\tau_2$.
Under these conditions only the fifth term of Eq.~(\ref{ECHO}), the so-called echo term, survives and it is independent of $\tau=\tau_1+\tau_2, (\tau_1=\tau_2)$ \cite{PCM}.
By using phase modulated pulses we can control the rotation axis of the Bloch vector~\cite{PHASE} for all three pulses of the echo sequence \cite{8PHASE}.
The phase cycling method makes it possible to extract only the echo term.
We are able to obtain stable echo decay curve during long time experiment.

\begin{figure}[t]
\begin{center}
\includegraphics[width=0.9\columnwidth]{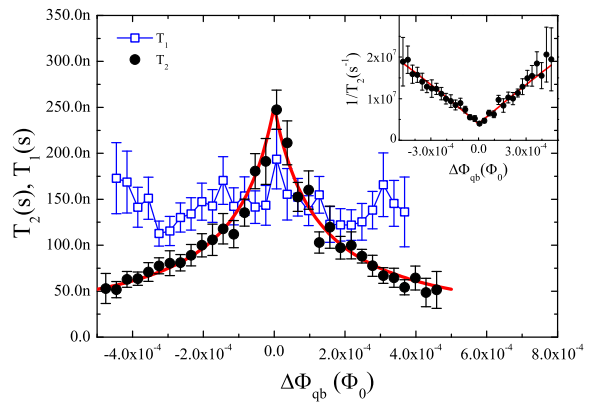}
\caption{(color online). Magnetic flux dependence of $T_1$ (open squares) and $T_2^{\rm echo}$ (filled circles).  
Although there is considerable scatter in the $T_1$ data, obviously there is only a weak flux dependence.
The largest $T_2^{\rm echo}$ values are obtained at the optimal point.
Since the $T_2^{\rm echo}$ data have been obtained
via the phase cycling technique, the difference of the magnitude of error bars of $T_1$ and $T_2^{\rm echo}$ is due to differences in the averaging number (each $T_2^{\rm echo}$ value is obtained from roughly eight times as many data as the $T_1$ value)~\cite{8PHASE}.
}
\label{T1T2}
\end{center}
\end{figure}

Figure.~\ref{T1T2} shows the magnetic flux dependence of $T_1$ and $T_2^{\rm echo}$ in the vicinity of the optimal point.
The $T_1$ values show some scatter but the flux dependence of $T_1$ is weak.
On the other hand, the $T_2^{\rm echo}$ values show a pronounced flux dependence.
In particular, there is 
a sharp maximum at the optimal point.
This tendency was also reported for the edge shared SQUID-flux qubit by Bertet et al.~\cite{PHOTONNOISE}.
At the optimal point, we have $T_2^{\rm echo}>T_1$.

It is well known~\cite{BLOCH} that the energy relaxation contributes to the dephasing process via
\begin{equation}
\label{BLOCH_WHITE}
\left(T_2^{\rm echo}\right)^{-1} = \left(2T_1\right)^{-1}+\Gamma_{\varphi}\ .
\end{equation}
Here $\Gamma_{\varphi}$ is the 
pure
dephasing rate due to the fluctuations 
in the energy splitting of the qubit.
The dephasing rate $\Gamma_{\varphi}$ 
consists of
 two contributions
$\Gamma_{\varphi}=\Gamma^0_{\varphi}+\Gamma^{\Phi}_{\varphi}$.
$\Gamma^0_{\varphi}$ should be identified with the non-magnetic dephasing processes or with the second-order contribution \cite{QUADRATIC}.
$\Gamma^{\Phi}_{\varphi}$ is the first-order contribution due to magnetic dephasing processes.
At the optimal point, the magnetic flux derivative of the energy dispersion is flat $\left|{\partial \Delta E \over \partial \Phi}\right|=0$, and there is no contribution of first-order magnetic flux fluctuations to the dephasing rate, i.e. $\Gamma^{\Phi}_{\varphi}=0$.
We obtain $T_2^{\rm echo} = 250$ ns at the optimal point and the average value of $T_1$ around the optimal point is about 140 ns.
From these values, we estimate the pure dephasing rate at the optimal point to be $\Gamma^0_{\varphi}= \left(2.3 \mu{\rm s}\right)^{-1}$.
{\it Remarkably, the dephasing time of the qubit $(T_2^{\rm echo})$ is almost entirely determined by the energy relaxation process, i.e. $T_2^{\rm echo} \approx 2 T_1$.}
In other words, the dephasing of the flux qubit  near the optimal point is predominantly determined by the high-frequency noise 
$S_{\Phi}\left(\omega\approx \Delta/\hbar \right)$.

For dephasing dominated by magnetic flux noise with a smooth spectrum near $\omega = 0$ this rate is given by
\begin{equation}
\label{GAMMA_1}
\Gamma^{\Phi}_{\varphi}=\frac{1}{2}
\left|{\partial \Delta E \over \partial
\Phi_x}\right|^2\,S_{\Phi}(\omega=0)\ .
\end{equation}
Here $\Delta E\equiv \Delta E(\Phi_x)$ is
the external, flux dependent qubit energy splitting and $S_{\Phi}(\omega=0)$ is the low frequency component of a correlation function of flux fluctuations.
From $\Delta E =
\sqrt{\epsilon^2(\Phi_x) + \Delta^2}$ we obtain $\partial \Delta
E/\partial \Phi_x=\cos\eta\,(\partial \epsilon/\partial \Phi_x$),
where $\tan\eta = \Delta/\epsilon$. 
Since the variation of the bias with
external flux, $\partial \epsilon/\partial \Phi_x$, is usually almost
constant over a wide interval of the applied flux.
$\Gamma^{\Phi}_\varphi$ becomes proportional to $\cos^2\eta$.
Taking into account that, according to our experimental findings, the $T_1$ time is almost independent of the applied flux and 
that a relation $\cos{\eta} \propto \epsilon$ is satisfied near the optimal point, a parabolic behavior of $1/T_2^{\rm echo}$ is expected in this region.
However, as shown in the 
inset of Fig.~\ref{T1T2} 
the measured $1/T_2^{\rm echo}$ versus applied flux curve is linear.
This result is very different from the prediction of Eq.~(\ref{GAMMA_1}).
As a consequence, our experimental data cannot be explained with a simple 
Bloch-Redfield type of decoherence theory
where
 a 
noise with short correlation time (white noise near $\omega=0$) is assumed.
In order to explain the observed behavior 
we need to consider decoherence from a long correlated 
noise.

Even when the 
pure
 dephasing is non-exponential 
the echo decay curve can be represented as the product of energy decay ($T_1$) and pure phase decay.
In this case Eq.~(\ref{BLOCH_WHITE}) is 
replaced 
by a non-exponential decay curve for the expectation value of 
$\sigma_z$ at the end of the echo sequence $\rho(t)\equiv\langle\sigma_z\rangle
= \exp{\left(-{t\over 2 T_1}\right)}\cdot \rho_{\rm echo}\left(t
\right)$ , where
$\rho_{\rm echo}\left(t\right)=\left<
\exp{\left\{ -i \left(\frac{\partial \epsilon}{\partial
\Phi}\right)\,\cos{\eta} \left[ \int_0^{t\over 2}
\Phi\left(\tau\right)d \tau-\int_{t\over 2}^{t}
\Phi\left(\tau\right) d \tau \right]\right\}}\right>$. 
Assuming
Gaussian flux fluctuations this gives
$\rho_{\rm echo}\left(t\right) =
\exp{\left[-{1\over 2}\left(\frac{\partial \epsilon}{\partial
\Phi}\right)^2 \cos^2{\eta}\int \frac{d
\omega}{2\pi}\,S_\Phi\left(\omega\right)  \frac{\sin^4{\omega t
\over 4}}{\left({\omega\over 4}\right)^2}\right]}$. 
In the case of white noise, $S_\Phi\left(\omega\right)={\rm const.}$, $\rho_{\rm echo}\left(t\right)$ gives a simple exponential decay curve, and the flux dependence of the decay rate follows Eq.~(\ref{GAMMA_1}).
On the other hand, for
$S_\Phi\left(\omega\right)={A\over f}={2\pi A\over \omega}$ ($1/f$-noise ~\cite{FINV})
$\rho_{\rm echo}\left(t\right) =
\exp{\left[-t^2\cdot\left(\frac{\partial \epsilon}{\partial
\Phi}\right)^2\cdot \cos^2{\eta}\cdot A\cdot\ln 2\right]}$.
This
decay law is 
Gaussian
 and the relaxation rate is proportional to $\left|{\partial \epsilon \over \partial \Phi} \cos{\eta}\right|$.

\begin{figure}[b]
\begin{center}
\includegraphics[width=0.75\columnwidth]{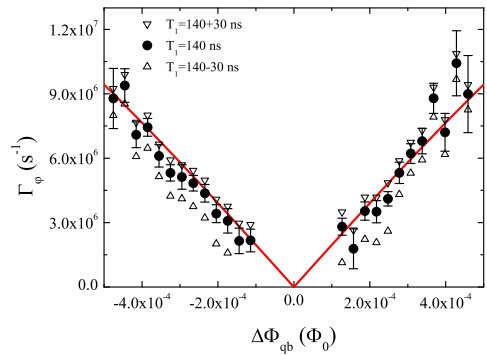}
\caption{
(color online). Magnetic flux dependence of $\Gamma_{\varphi}$ values.
The $\Gamma_{\varphi}$ values are derived from fits of the data by the $1/f$ fluctuation model using a constant $T_1$ value. 
The filled circles represent $\Gamma_{\varphi}$ values estimated using an average $T_1$ value of 140 ns. 
Since the $T_1$ values have some scattering, we also estimated $\Gamma_{\varphi}$ values using both $T_1 = 140 \pm 30$ ns. The triangles and inverted triangles represent the $\Gamma_{\varphi}$ values estimated from $T_1 = 110$ ns and 170 ns, respectively. The red solid line shows a linear fit to the $\Gamma_{\varphi}$ data.
}
\label{T2INV}
\end{center}
\end{figure}

We do not attempt to determine the echo decay shape from the observed relaxation curve
because
our experimental data do not have precision enough to distinguish between the different types of decay curves (e.g. simple exponential, Gaussian or algebraic). Instead, we assume a special decay shape based on the $1/f$ fluctuations spectrum 
$
\rho_{}\left(t\right)=\exp{\left(-{t\over 2 T_1}\right)}\cdot 
\exp{\left(-\Gamma_{\varphi}^0 t\right)}\cdot \exp{\left(-{\Gamma^\Phi_{\varphi}}^2 t^2\right)}
$
. And we find the magnetic flux dependence of $\Gamma^\Phi_{\varphi}$ by fitting 
of this formula.
In this fitting $T_1$ and $\Gamma^0_{\varphi}$ 
are fixed and $\Gamma^{\Phi}_{\varphi}$ is the fitting parameter.
This relaxation function 
 is constructed from three components. 
The first one describes the energy relaxation ($T_1$ process). In the second component 
$\Gamma^0_{\varphi}$  is the phase relaxation rate, which is caused by non-magnetic fluctuations \cite{IBIAS} 
and second-order contributions of magnetic fluctuations. 
Near the optimal point, the second-order $\Phi_x$ contributions of $\Gamma^0_{\varphi}$ is small then we can neglect it.
Finally, $\Gamma^\Phi_{\varphi}$ is the phase relaxation rate due to $1/f$ magnetic fluctuations.
We obtain $T_1=140$ ns from the energy relaxation curve and $\Gamma_{\varphi}^0=(2.3$ $\mu{\rm s})^{-1}$ from the echo decay rate at the optimal point.
Figure.~\ref{T2INV} shows the magnetic flux dependence of the resulting $\Gamma^\Phi_{\varphi}$.
Very close to the optimal point, the contribution of $\Gamma^\Phi_{\varphi}$ to the echo decay curve is very small compared to that of $1/T_1$.
In this region the
$\Gamma^\Phi_{\varphi}$ values 
cannot be extracted with sufficient accuracy 
from the echo decay curve.
Apart from this region, the magnetic flux dependence of $\Gamma^\Phi_{\varphi}$ is well fitted by expression 
$\Gamma^\Phi_{\varphi} \propto \left|{\partial \Delta E\over \partial \Phi}\right|=\left|{\partial \epsilon \over \partial \Phi} \cos{\eta}\right|$.
This means that $\Gamma^\Phi_{\varphi}$ is proportional to absolute value of the magnetic flux difference from the optimal point.
Thus our experimental result 
suggests the existence of $1/f$ type magnetic fluctuations in the frequency range of $\omega \approx O\left(\Gamma^{\Phi}_\varphi\right)$.
The magnitude of the $1/f$ flux noise can be estimated 
for 
the $A$ value of the correlation function 
$S_\Phi\left(\omega\right)={2\pi A\over\omega}$.
From the spectrum dispersion and the magnetic flux dependence of $\Gamma^\Phi_{\varphi}$
we obtain 
 $A\approx\left(10^{-6} \Phi_0\right)^2$.
This value is of the same order as for other superconducting circuit \cite{AVAL1}.

In summary, we have performed a systematic study of the phase and energy 
relaxation of a superconducting flux qubit by measuring the magnetic flux dependence of the energy relaxation ($T_1$) time 
and of the echo decay time ($T_2^{\rm echo}$). 
At the optimal point
we find that the contribution of magnetic fluctuations is negligible.
The $T_2^{\rm echo}$ time of the qubit is then determined by the energy relaxation resulting in
$T_2^{\rm echo}\approx 2T_1$.
Thus, in order to improve the coherence time of our flux qubit at the optimal point, {\it we need to suppress high frequency transversal fluctuations} $S_{x,y}\left(\Delta/\hbar\right)$.
Away from the optimal point the magnetic flux fluctuations dominate the echo dephasing time $T_2^{\rm echo}$.
The observed linear dependence of $\Gamma^\Phi_{\varphi}$ on magnetic flux around the optimal point suggests that {\it the magnetic fluctuations have a $1/f$-type spectrum in a frequency range of the order of $1/T_2^{\rm echo}$.}
To our knowledge these results provide the first experimental 
clue for
the possible origin of decoherence in a superconducting flux qubit, which is not sharing an edge with its dc SQUID detector.

We would like to thank M.Ueda and R.Gross for fruitful discussions.
This work is partly supported by 
CREST-JST and 
JSPS-KAKENHI(18201018).
One of us, F.D., acknowledges support from the SFB 631 of the Deutsche Forschungsgesellschaft.
A.S. acknowledges support from IST FP6-015708 EuroSQIP. 

Note added: After complete this work, we became aware of similar results by F. Yoshihara {\it et al.} \cite{NEC}.


\begin{thebibliography}{9}
\bibitem{PCM} 
If the echo conditions are satisfied only approximately, e.g. because of non-perfect $\pi/2$ or $\pi$ pulses, other terms may contribute. The fact that these terms depend on $\delta\omega\tau$ explains some of the beatings in the echo decay curve observed sometimes in our experiments.
\bibitem{8PHASE} We have four choices of the second pulse phase ($0^{\circ}$, $90^{\circ}$, $180^{\circ}$, $270^{\circ}$) and two for the first pulse phase ($0^{\circ}$, $180^{\circ}$) as a pulse sequence. There are eight combinations of these pulse phases. By the sign of the fifth term of Eq.~(\ref{ECHO}), we add or subtract echo decay data of each pulse combination.
\bibitem{IBIAS} $\Gamma^0_{\varphi}$ may be caused by SQUID bias current fluctuations. But contributions of bias current fluctuations are weak. The coupling between our sample's qubit and SQUID is weak, because our qubit sample does not share 
any edges of 
qubit and SQUID.
\end{thebibliography}
\end{document}